\documentclass[5p,twocolumn,times,number]{elsarticle}

\usepackage{graphicx}
\usepackage{amsmath} 

\usepackage{lineno}

\begin{document}

\begin{frontmatter}

\title{The MPGD-Based Photon Detectors for the upgrade of 
COMPASS RICH-1 and beyond}

\author[add-ts-infn]{J.~Agarwala}
\author[addto-infn]{M.~Alexeev}
\author[add-aveiro]{C.D.R.~Azevedo}
\author[add-trieste]{F.~Bradamante}
\author[add-trieste]{A.~Bressan}
\author[add-fr]{M.~B\"uchele}
\author[add-to]{M.~Chiosso}
\author[add-trieste]{C.~Chatterjee}
\author[add-trieste]{P.~Ciliberti}
\author[add-ts-infn]{S.~Dalla~Torre\corref{cor}}
\ead{silvia.dallatorre@ts.infn.it}
\author[add-ts-infn]{S.~Dasgupta}
\author[addto-infn]{O.~Denisov}
\author[add-prague]{M.~Finger}
\author[add-prague]{M.~Finger~Jr}
\author[add-fr]{H.~Fischer}
\author[add-ts-infn]{M.~Gregori}
\author[add-ts-infn]{G.~Hamar}
\author[add-fr]{F.~Herrmann}
\author[add-ts-infn]{S.~Levorato}
\author[add-trieste]{A.~Martin}
\author[add-ts-infn]{G.~Menon}
\author[add-aless]{D.~Panzieri}
\author[add-trieste]{G.~Sbrizzai}
\author[add-fr]{S. Schopferer}
\author[add-prague]{M. Slunecka}
\author[add-liberec]{M. Sulc}
\author[add-ts-infn]{F.~Tessarotto}
\author[add-aveiro]{J.F.C.A.~Veloso}
\author[add-ts-infn]{Y.~Zhao}

\cortext[cor]{Corresponding author}

\address[add-ts-infn]{INFN Trieste, Trieste, Italy}
\address[addto-infn]{INFN Torino, Torino, Italy}
\address[add-aveiro]{University of Aveiro, Aveiro, Portugal}
\address[add-trieste]{University of Trieste and INFN Trieste, Trieste, Italy}
\address[add-fr]{Universit\"at Freiburg, Freiburg, Germany}
\address[add-to]{University of Torino and INFN Torino, Torino, Italy}
\address[add-prague]{Charles
University, Prague, Czech Republic and JINR, Dubna, Russia}
\address[add-aless]{University of East
Piemonte, Alessandria and INFN Torino, Torino, Italy}
\address[add-liberec]{Technical University of Liberec, Liberec, Czech Republic}

\begin{abstract}
After pioneering gaseous detectors of single photon for RICH 
applications using CsI solid state photocathodes in MWPCs within 
the RD26 collaboration and by the constructions for the RICH detector 
of the COMPASS experiment at CERN SPS, in 2016 we have upgraded 
COMPASS RICH by novel gaseous photon detectors based on MPGD 
technology. Four novel photon detectors, covering 
a total active area 
of 1.5~m$^2$, have been installed in order to cope with the 
challenging efficiency and stability requirements of the COMPASS
physics programme. These detectors are the first application in an 
experiment of MPGD-based single photon detectors.
All aspects of the upgrade are presented, including engineering, 
mass production, quality assessment and performance.
\par
Perspectives for further developments in the field of gaseous 
single photon detectors are also presented.

\end{abstract}

\begin{keyword}
Gaseous photon detectors \sep MPGD 
\sep Thick GEM \sep resistive MICROMEGAS \sep COMPASS

\end{keyword}

\end{frontmatter}

\section{Introduction}
THE RICH-1 detector~\cite{rich-1} of the 
COMPASS Experiment~\cite{compass}
at CERN SPS has been
upgraded: four new Photon Detectors 
(unit size: 600$\times$600~mm$^2$),
based on MPGD technology and covering a total active area
of 1.5~m$^2$ replace the previously used MWPC-based photon
detectors in order to cope with the challenging efficiency and
stability requirements of the new COMPASS measurements.
In COMPASS RICH-1, MPGD photon detectors are used for
the first time in a running experiment. This realization 
also opens the way of a more extended
use of novel gaseous photon detectors in the domain of the
Cherenkov imaging technique for Particle IDentification (PID),
key detectors in several research sectors and, in particular, in
hadron physics. The relevance is related to the role of gaseous
photon detectors, which are still the only available option to
instrument detection surfaces when insensitivity to magnetic
field, low material budget, and affordable costs in view of
large detection surfaces are required. The MPGD-based photon
detectors overcome the limitation of the previous generation of
gaseous photon detector thanks to two essential performance
characteristics: reduced ion and photon backflow to the photocathode,
namely reduced ageing and increased electrical stability, and
faster signal development, namely higher rate capabilities.
\section{The novel photon detectors}
\begin{figure}
\centering
\includegraphics[width=0.99\linewidth]{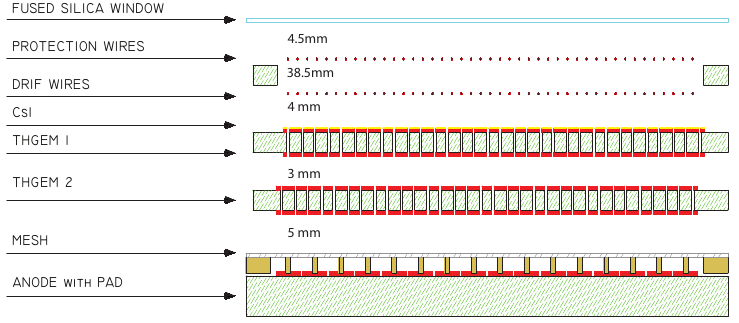}
\caption{Sketch of the hybrid single photon detector: 
two staggered THGEM
layers are coupled to a resistive bulk MM. Image not to scale.}
\label{fig:architecture}
\end{figure}

The detector architecture is the result of a 
seven-year R\&D activity~\cite{rd}. It is based on a  hybrid
MPGD combination (Fig.~\ref{fig:architecture}), consisting in
two layers of THick GEMs
(THGEM)~\cite{thgem} followed by a resistive 
MicroMegas (MM)~\cite{mm} on a pad segmented
anode. The first THGEM 
also acts as a reflective
photocathode: its top face is coated with a CsI film.
The feedback of photons generating in the multiplication 
process is suppressed by the presence of two THGEM layers, 
while the large majority of the ions from multiplication
are trapped in the MM stage. MPGD properties ensure
signal development in about 100 ns. 
\par 
Each of the four large  (600$\times$600~mm$^2$) single photon
detectors is formed by two identical modules 600$\times$300~mm$^2$,
arranged side by side. The THGEM geometrical parameters are: 
470~$\mu$m thickness,
400~$\mu$m hole diameter and 800~$\mu$m pitch. Holes are rim-less, 
namely there is no uncoated area
around the hole edge. In order to mitigate the effect of 
occasional discharges, the top and bottom electrodes 
of each THGEM are
segmented in 12 parallel areas separated by 0.7~mm clearance, 
each biased via an individual protection resistor. Therefore, 
discharges only affect a single sector and the operating conditions 
are restored in about 10~s.
The two layers are 
staggered, namely there is complete misalignment between 
the two set of holes: it is so possible to enlarge the electron 
cloud reaching the MM stage, therefore favoring larger gain
in the last amplification stage.
\par
\begin{figure}
\centering
\includegraphics[width=0.99\linewidth]{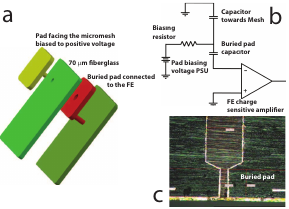}
\caption{a) Sketch of the capacitive coupled readout pad. 
The biasing voltage is
distributed via independent 470~M$\Omega$
resistors to the pad facing the micromesh
structure. The buried pad is isolated via 70~$\mu$m 
thick fiberglass and connected
to the front end chip. 
b) Schematic of the capacitive coupled pad principle
illustrated via discrete element blocks. 
c) Metallography section of the PCB:
detail of the through-via connecting the 
external pad through the hole of the
buried pad.}
\label{fig:resistive-mm}
\end{figure}

The MMs have a 
gap of 128~$\mu$m;
they are built using MM bulk technology~\cite{bulk} 
using 300~$\mu$m
diameter pillars with 2~mm pitch. Their resistivity
is by an original implementation 
(Fig.~\ref{fig:resistive-mm}). The
MM anode is segmented in 7.5$\times$7.5~mm$^2$ pads.  
The 0.5~mm clearance
between pads prevents the occasional discharges to
propagate towards the surrounding pads: the voltage drop of
the anodic pads surrounding a tripping one is about 2 V over
the typical 600 V operation voltage, 
causing a local gain drop lower
than 4\%. The nominal voltage condition of the pad where 
the discharge occurred
is restored in about 1~s.
The detector is operated with
Ar:CH$_4$~=~50:50 gas mixture, which ensures effective extraction of photoelectrons from the photocathode.   
\section{Construction, quality control of the components, assembly and installation}
\begin{figure}
\centering
\includegraphics[width=0.80\linewidth]{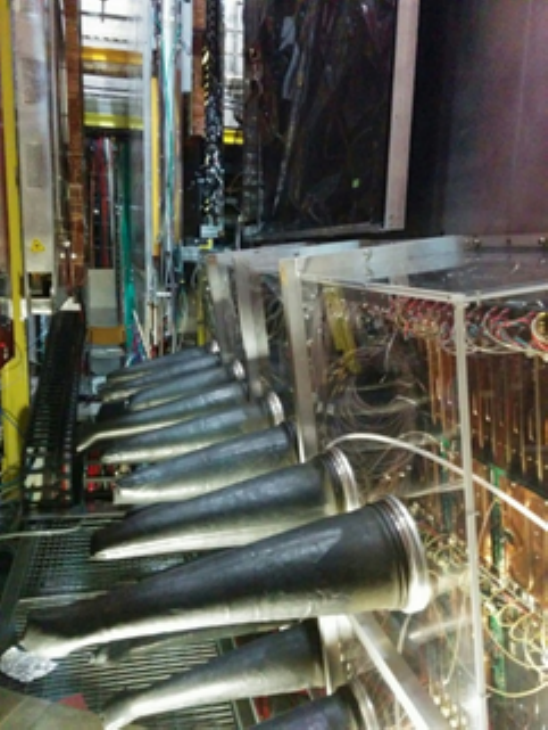}
\caption{Picture of the glove 
box matching the RICH vessel mechanics used for 
the photon detector installation on the RICH.}
\label{fig:glovebox}
\end{figure}

The electrical stability of large-size THGEMs is a critical issue.
A dedicated protocol  has been elaborated for finishing the industrial produced THGEMs . It includes polishing 
with fine grain pumice powder, cleaning with water at high pressure, ultrasonic bath with
Sonica PCB solution (PH11), rinsing with distilled water
and backing in oven at 160$^o$C. The procedure moves 
THGEM breakdown voltage to at least 90\% of 
the phenomenological
Paschen limit~\cite{paschen}.
The quality control of the detector components includes:
\begin{itemize} 
\item
the preselection of the raw material for the PCB that will
form the THGEMs in order to use only foils with
homogeneous thickness to guarantee the homogeneity of
the gain;
\item 
the THGEM control by optical inspection, by collecting and
analyzing microscope images scanning by samples the
large multiplier surface;
\item
the THGEM validation by gain maps using the multipliers in
single layer detectors; gain uniformity at 7\% r.m.s. is obtained;
\item
the collection of MM gain maps illuminating the
detectors by an X-ray gun station; gain uniformity 
at 5\% r.m.s. is obtained;
\item
the measurement of the quantum efficiency of the CsI 
photocathodes, which is performed immediately after 
the coating process; the uniformity within a photocathode is at 
the 3\% level r.m.s. and among the whole production at 
the 10\% level r.m.s.;
\item
the gas leak checks and overall electrical stability checks of
the final detectors.
\end{itemize}
The presence of CsI photocathodes, that must never be 
exposed to air to fully preserve their quantum efficiency, 
impose to perform
detector assembly,  transportation and
installation in  glove boxes flushed with N$_2$. 
Figure~\ref{fig:glovebox} presents the dedicated glove 
box matching the RICH vessel mechanics used for 
the on-detector installation.
\section{The high voltage system}
An essential tool for the detector commissioning is the High
Voltage (HV) control system, which also grants the voltage and
current monitoring and data logging. The power supplies are
commercial ones by CAEN inserted in a SY4527 mainframe.
The four detectors are organized, from the HV supply point of
view, in four independent sectors each; nine different electrode
types, each one with its specific role, are present in the
multilayer detectors. The total number of HV channels is 136. 
Manual setting and control of all these
HV channels would be both unpractical and unsafe.
The voltages and currents of all the channels are read-out
and recorded at 1~Hz frequency. If the current spark rate is
above a given value the voltage is automatically readjusted.
The system also provides automatic voltage adjustment to 
compensate for the variation of the environmental parameters, 
namely pressure and temperature, that can affect the detector gain. 
Gain stability at 
the 5\% level over months of operation has been obtained.
\section{Preliminary performance results}
No HV trip is observed during detector operation: 
thanks to the resistors protecting the THGEM segments and 
the MM pads, in case of occasional discharges, 
only current sparks 
are observed, which temporary affect the local performance. 
The sparks 
in the two THGEM layers are fully correlated. The sparks 
observed in the MM are induced by the THGEM sparks. The 
restoration after a current spark 
is completed within 10~s and the current spark rate is 
typically 1/h/detector (600$\times$600~mm$^2$). These figures result in totally negligible dead-time related to sparks.
\par
The measured rate of ion backflow to the photocathode is at the 3\% level.
The electronics noise, substantially uniform over the detector surface, is 
at the 900 electrons equivalent level (r.m.s).
\par
\begin{figure}
\centering
\includegraphics[width=0.99\linewidth]{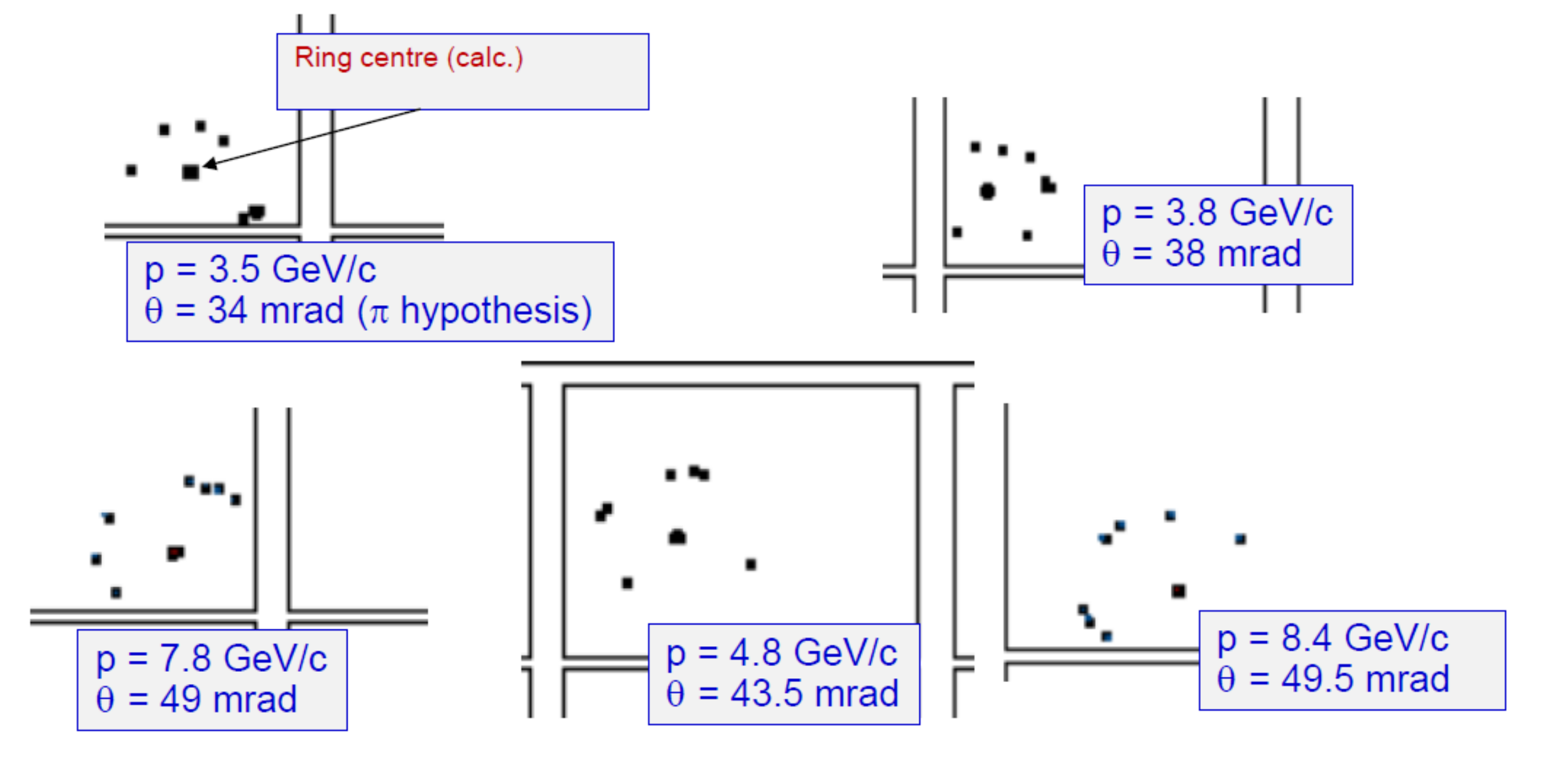}
\caption{Images of hit pattern in the novel photon detectors.
The center of the expected ring patterns is obtained 
from the reconstructed particle trajectories; the particle momentum 
and the expected Cherenkov angle in the pion hypothesis are also reported. 
No image elaboration or background subtraction is applied.}
\label{fig:rings}
\end{figure}
\begin{figure}
\centering
\includegraphics[width=0.99\linewidth]{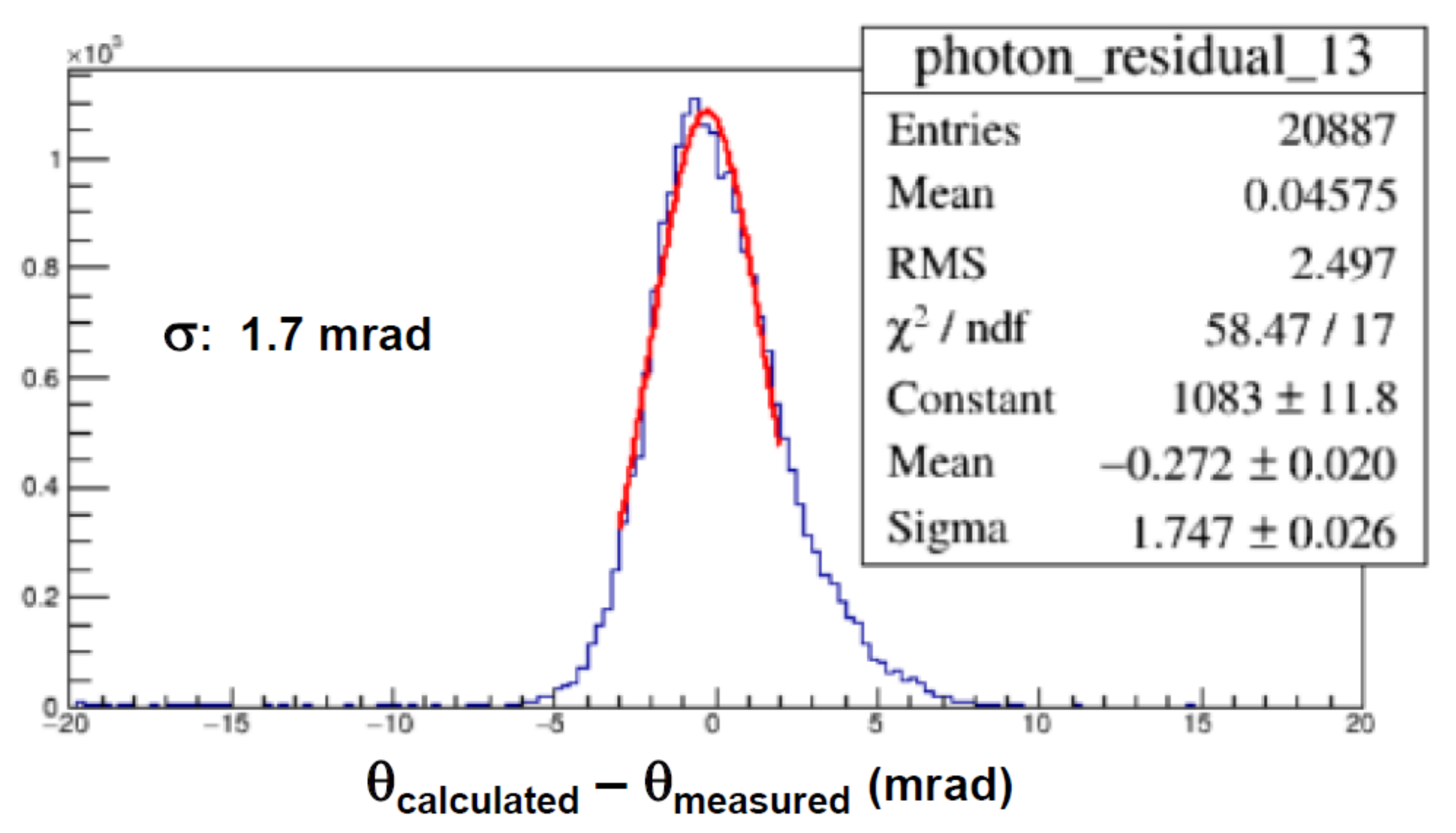}
\caption{Distribution of the difference between the Cherenkov angle
calculated from the reconstructed particle momentum and the 
Cherenkov angle provided by single detected photoelectrons;  a sample of identified pions is used.}
\label{fig:resolution}
\end{figure}
\begin{figure}
\centering
\includegraphics[width=0.99\linewidth]{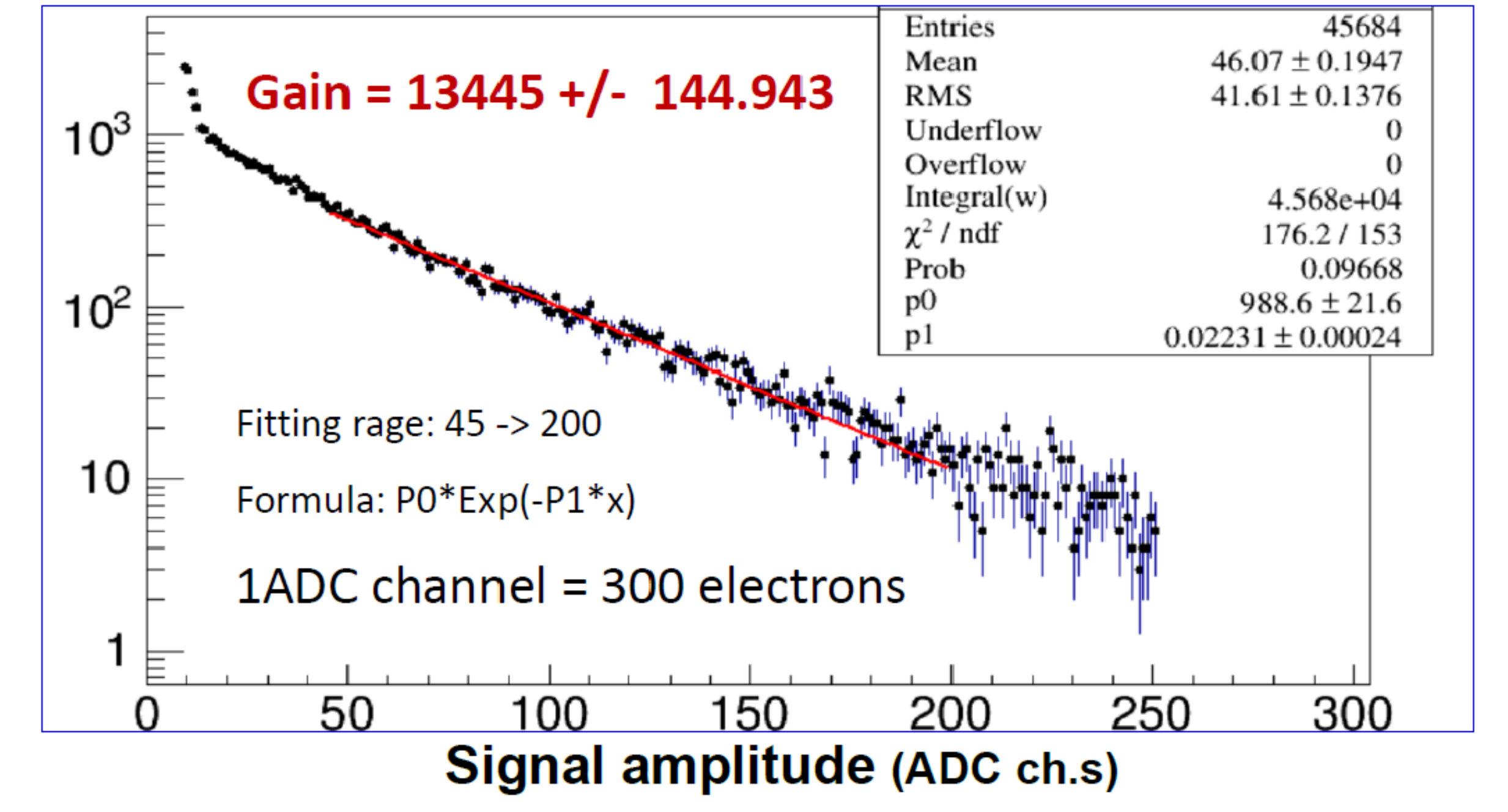}
\caption{Amplitude distribution for a sample of hits contributing 
to ring Cherenkov images.}
\label{fig:gain}
\end{figure}

The images generated in the photon detectors are clean and 
affected by very limited background (Fig.~\ref{fig:rings}).
The detector resolution in the measurement of the Cherenkov angle 
from single photoelectrons is  1.7-1.8~mrad r.m.s., fully
matching the expectation (Fig.~\ref{fig:resolution}). 
The amplitude 
spectrum of the photoelectron signals is expected 
to be exponential.
This is verified for pure photoelectron samples, 
obtained selecting 
hits contributing to ring images: the exponential behavior 
is present over more than two orders of
magnitude (Fig.~\ref{fig:gain}).
The detector gain is extracted from a fit of the spectrum 
and it ranges between 13k and 14k. The efficiency for single 
photoelectron detection is obtained from the gain and the 
threshold applied to the 
signal and it results higher than 80\%. The noise contributing 
to the ring images can be estimated from the spectrum deviation 
from a pure exponential function at small amplitude and it 
is at the 10\% level. A preliminary estimate of the number 
of detected photoelectrons 
per particle extrapolated to the saturation angle indicates 
11 detected photoelectrons.
\par
The high effective 
gain, the gain stability and the number of detected 
photoelectrons per ring
satisfy all the prerequisite requirements to ensure effective hadron
identification and stable performance
with the novel RICH-1 photon detectors.
\section{Future perspective}
The future Electron-Ion Collider (EIC)~\cite{eic} requires 
hadron identification at high momenta, a mission that can 
only be accomplished by RICH counters with extended gas radiator.
The use of RICHes in the setup of collider experiments implies 
specific challenges. The radiator cannot be too extended to 
limit the overall apparatus size, imposing the need 
to detect more photoelectrons per radiator unit length. 
The photon detectors have to operate in presence of magnetic field.
A recent test-beam exercise has demonstrated the possibility 
to increase the number of detected photoelectrons by selecting 
the far UV range around 120~nm~\cite{windowless}. For this purpose, the 
RICH prototype has been operated window-less and CF$_4$ has been used at 
the same time as radiator gas and detector gas.
Therefore, we have started an R\&D program to 
match these specific requirements. It includes 
the exploratory study of a new option for the photoconverter: 
coating by hydrogenized nanodiamond powder~\cite{nd}.
\section{Conclusion}
The preliminary results obtained in  the characterization of 
novel MPGD-based photon detectors, in particular high effective 
gain, gain stability and number of photons per ring, indicate that they 
will fully accomplish the mission of increasing the 
stability and efficiency
of the photon detector system of COMPASS RICH-1.
They also represent a
technological achievement. In fact, for the first time in a 
running experiment, THGEMs are successfully used, 
single photon detection is accomplished by MPGDs, 
MPGDs are operated at gains larger than 10k.
\par
We have offered indications that MPGD-based photon detectors 
have a mission to accomplish also in the future, in particular in the hadron physics sector. 
\section*{Acknowledgments}
The authors are grateful to the colleagues of the 
COMPASS Collaboration for continuous support and encouragement.
\par
This work is partially supported by the 
H2020 project AIDA-2020, GA no. 654168.
\par 
J. Agarwala is supported by an ICTP TRIL fellowship.



\begin{thebibliography}{10}
\expandafter\ifx\csname url\endcsname\relax
  \def\url#1{\texttt{#1}}\fi
\expandafter\ifx\csname urlprefix\endcsname\relax\def\urlprefix{URL }\fi

\bibitem{rich-1}
E.~Albrecht, et al., Status and characterisation of COMPASS RICH-1, 
Nucl. Instrum. Meth. A553 (2005) 215; 
P.~Abbon
et al., Read-out electronics for fast photon detection with 
COMPASS RICH-1, Nucl. Instrum. Meth. A587 (2008) 371; 
P.~Abbon et al.,   Design and construction of the fast photon 
detection system for COMPASS RICH-1, Nucl.
Instrum. Meth. A616 (2010) 21; 
P.~Abbon et al., Particle identification with COMPASS RICH-1,
Nucl. Instr. and Meth.A631 (2011) 26.

\bibitem{compass}
The COMPASS Collaboration, P.~Abbon et al., 
The COMPASS experiment at CERN, Nucl. Instrum. Meth. A577 (2007) 455; 
the COMPASS Collaboration, P.~Abbon et al., The COMPASS setup for 
physics with hadron beams, 
Nucl. Instrum. Meth. A779 (2015) 69.

\bibitem{rd}
M.~Alexeev et al., The quest for a third generation of gaseous 
photon detectors for Cherenkov imaging counters, 
Nucl. Instrum. Meth. A610 (2009) 174; 
M.~Alexeev et al., THGEM based photon detector for 
Cherenkov imaging applications, Nucl. Instrum. Meth. A617 (2010) 396;  
M.~Alexeev et al.,     Micropattern gaseous photon detectors for 
Cherenkov imaging counters, Nucl. Instrum. Meth. A623 (2010) 129; 
M.~Alexeev et al., Development of THGEM-based photon detectors for 
Cherenkov Imaging Counters, 2010 JINST 5 P03009; 
M.~Alexeev et al., Progress towards a THGEM-based detector of 
single photons, Nucl. Instrum. Meth. A639 (2011) 130;
M.~Alexeev et al., Detection of single photons with 
ThickGEM-based
counters, 2012 JINST 7 C02014; 
M.~Alexeev et al., Detection of single photons with 
THickGEM-based counters, Nucl. Instrum. Meth. A695 (2012) 159;
M.~Alexeev et al.,  Development of THGEM-based Photon Detectors for 
COMPASS RICH-1, Physics Procedia 37 (2012) 781; 
M.~Alexeev et al., THGEM-based photon detectors for the upgrade of 
COMPASS RICH-1, Nucl. Instrum. Meth. A732 (2013) 264; 
M.~Alexeev et al., Ion backflow in thick GEM-based detectors of 
single photons, 2013 JINST 8 P01021; 
M.~Alexeev et al., Status and progress of novel photon 
detectors based on THGEM and hybrid MPGD architectures, 
2013 JINST 8 C12005; 
M.~Alexeev et al., Progresses in the production of large-size 
THGEM boards,
2014 JINST 9 C03046; 
M.~Alexeev et al., MPGD-based counters of single photons
developed for COMPASS RICH-1, 2014 JINST 9 C09017; 
M.~Alexeev et al., Status and progress of the novel photon 
detectors based on THGEM and hybrid MPGD architectures, 
Nucl. Instrum. Meth. A766 (2014) 133; 
M.~Alexeev et al., 
MPGD-based counters of single photons for
Cherenkov imaging counters, PoS (TIPP2014) 075;
M.~Alexeev et al., The gain in Thick GEM multipliers and 
its time-evolution, 2015 JINST 10 P03026;  
M.~Alexeev et al., Status of the development of large area 
photon detectors based on THGEMs and hybrid MPGD architectures 
for Cherenkov imaging applications, 
Nucl. Instrum. Meth. A824 (2016) 139.

\bibitem{thgem}
L.~Periale et al., Detection of the primary scintillation light 
from dense Ar, Kr and Xe with novel photosensitive gaseous detectors, 
Nucl. Instrum. Meth. A478 (2002) 377; 
P.~Jeanneret, Time Projection Chambers and detection of neutrinos,
PhD thesis, Neuchatel University, 2001; 
P.S.~Barbeau et al, Toward coherent neutrino detection using 
low-background micropattern gas detectors
IEEE NS-50 (2003) 1285; 
R.~Chechik et al,  Thick GEM-like hole multipliers: properties 
and possible applications, Nucl. Instrum. Meth. A535 (2004) 303.

\bibitem{mm}
Y.~Giomataris et al., MICROMEGAS: a high-granularity 
position-sensitive gaseous detector for high particle-flux 
environments, Nucl. Instrum. Meth. A376 (1996) 29.

\bibitem{bulk}
I.~Giomataris et al., Micromegas in a bulk, 
Nucl. Instrum. Meth. A560 (2006) 405.

\bibitem{paschen}
F.~Paschen, \"Uber die zum Funken\"ubergang in Luft, Wasserstoff 
und Kohlens\"aure bei verschiedenen Drucken erforderliche 
Potentialdifferenz, Annalen der Physik 273 (1889) 69.

\bibitem{eic}
A.~Accardi et al., Electron-Ion Collider: The next QCD frontier,
Eur. Phys. J. A52 (2016) 268.

\bibitem{windowless}
M.~Blatnik et al., 
Performance of a Quintuple-GEM Based RICH Detector Prototype,
IEEE NS 62 (2015) 3256.

\bibitem{nd}
L.~Velardi, A.~Valentini, G.~Cicala, UV photocathodes based on
nanodiamond particles: Effect of carbon hybridization on the 
efficiency,  Diamond and Related Materials 76 (2017) 1.

\end{thebibliography}
\end{document}